# Effect of Site-disorder, Off-stoichiometry and Epitaxial Strain on the Optical Properties of Magnetoelectric Gallium Ferrite

Amritendu Roy[1], Somdutta Mukherjee[2], Surajit Sarkar[2], Sushil Auluck[3], Rajendra Prasad[2], Rajeev Gupta[2,4] and Ashish Garg[1*]

[1]Department of Materials Science & Engineering, Indian Institute of Technology, Kanpur-208016, India
[2]Department of Physics, Indian Institute of Technology, Kanpur - 208016, India
[3]National Physical Laboratory, Dr. K. S. Krishnan Marg, New Delhi-110012, India
[4]Materials Science Programme, Indian Institute of Technology, Kanpur - 208016, India

**ABSTRACT**

We present a combined experimental-theoretical study demonstrating the role of site disorder, off-stoichiometry and strain on the optical behavior of magnetoelectric gallium ferrite. Optical properties (band-gap, refractive indices and dielectric constants) were experimentally obtained by performing ellipsometric studies over the energy range 0.8 – 4.2 eV on pulsed laser deposited epitaxial thin films of stoichiometric gallium ferrite with *b*-axis orientation and the data was compared with theoretical results. Calculations on the ground state structure show that the optical activity in $GaFeO_3$ arises primarily from O2p-Fe3d transitions. Further, inclusion of site disorder and epitaxial strain in the ground state structure significantly improves the agreement between the theory and the room temperature experimental data substantiating the presence of site-disorder in the experimentally derived strained $GaFeO_3$ films at room temperature. We attribute the modification of the ground state optical behavior upon inclusion of site disorder to the corresponding changes in the electronic band structure, especially in Fe3d states leading to a lowered band-gap of the material.

---

[*] Corresponding author, Tel: +91-512-2597904; FAX - +91-512-2597505, E-mail: ashishg@iitk.ac.in

# I. Introduction

Optical properties of piezoelectric and ferroelectric oxides have been of interest for a variety of applications [1-4] ranging from optical waveguides, [2] photocatalysis, [5] and infrared detection [3, 4] to more recently photovoltaics. [6, 7] As an additional degree of freedom in many of these materials such as magnetoelectrics and multiferroics, magnetism often yields a reduced band gap [8, 9], enhancing their suitability for photovoltaic applications. Further, tunabilty of the band gap *vis-à-vis* electronic structure in these materials could be accomplished by tailoring the crystal structure where external perturbations such as doping [10] and strain [11] play crucial role in determining structural symmetry as well as physical properties *e.g.* optical properties. To fabricate efficient optical devices, it is essential to understand the microscopic effects of such perturbations *vis-à-vis* their contributions to the electronic structure as well as optical properties. In this regard, first-principles density functional theory (DFT) based studies have been quite successful in predicting and analyzing the ground state as well as electronic structure and optical properties of complex oxide systems. [12, 13] The disadvantage of underestimation of electronic band gap by conventional LDA and GGA functionals [14, 15] of DFT could be avoided by suitably scaling the calculated results with the experimental observations.

We have chosen to study the optical properties of gallium ferrite ($GaFeO_3$ or GFO) which is a prospective room temperature magnetoelectric with comparatively small band gap (~2.5-3 eV) [14, 16] and therefore, attractive for potential photovoltaic applications. Further, GFO exhibits a number of exciting optical phenomena such as optical magnetoelectric effect, [17, 18] and magnetization induced non-linear second harmonic Kerr effect. [19, 20] The observed ferrimagnetism, which can be tuned by tailoring the Ga:Fe ratio [21, 22] and processing conditions, [21-23] is believed to be the manifestation of inherent cation site disorder [21, 24] emanating from almost similar sizes of Fe and Ga. Previous optical studies on GFO single crystals [25] and thin films [16] using ellipsometry and absorption studies showed red-shift of the fundamental absorption edge with increasing Fe content. However, microscopic origin of such red-shift is rather generalized and one needs to decouple the possible contributions of external factors such as Fe content, cationic site disorder and epitaxial strain to completely understand the optical response of the material. In this context, systematic first-principles calculations along with appropriate experimental data could provide an atomistic insight into optical response of the material.

In this paper, we present the results of a combined experimental and theoretical study of the linear optical properties of GFO. Epitaxial thin films of GFO were chosen specifically to understand the combined effects of inherent cation site disorder as well as epitaxial strain on its optical properties which were decoupled using first-principles' calculations. Our calculations show that the inter-band transitions, responsible for optical activities of GFO are primarily due to O2p-Fe3d transitions. More importantly, we clearly observe that incorporation of cation site disorder and epitaxial strain into the ground state structure yields a much improved agreement between the theoretical predictions and experimental observations substantiating the role of cation site disorder which is explained in terms of modification of the Fe 3d bands. Subsequent parts of the paper are organized as follows: section II discusses the experimental techniques and calculation methodologies used in the study, section III presents the results of both ellipsometric measurements and first-principles calculations on the ground state structure along with external effects such as disorder, strain, off-stoichiometry and hydrostatic pressure and section IV summarizes the manuscript.

**II. Experimental and Calculation Details**

GFO thin films were grown using pulsed laser deposition technique on (100) oriented cubic yttria stabilized zirconia (YSZ) substrates from a stoichiometric target of GFO (Ga:Fe = 1). Film growth was carried out using KrF excimer laser (λ=248 nm) in an oxygen ambient (pO$_2$ ~ 0.53 mbar) at a substrate temperature of 800°C using a laser fluence of 2 J/cm$^2$ at a laser repetition rate of 3 Hz. As-grown films were subsequently cooled slowly to room temperature under the same ambient pressure. X-ray diffraction of the films was performed using a high resolution PANalytical X'Pert PRO MRD thin film diffractometer using CuKα radiation. Ellipsometric measurements were carried out using HORIBA JOBIN-YVON spectroscopic ellipsometer (SE) over the energy range of 0.8-4.2 eV with an incidence angle of 70°.

For calculations of the optical properties of the ground state structure of GFO, [14, 24] we employed density functional theory (DFT+U) [26] using pseudopotential based, Vienna *Ab-initio* simulation package (VASP) [27] and applied the projector augmented wave method (PAW) [28]. Kohn-Sham equation [29] was solved using the generalized gradient approximation (GGA+U) method (U = 5 eV, J = 1 eV) with the optimized version of Perdew-Burke-Ernzerhof functional for solids (PBEsol). [30] For calculations, we included three valence electrons of Ga ($4s^2 4p^1$), eight for Fe ($3d^7 4s^1$) and six for O ($2s^2 2p^4$) ions. A plane wave energy cut-off of 550 eV was used. We used Monkhorst-Pack [31] 4×4×4 mesh in our calculations. To check the robustness of our calculations, we also repeated some of our calculations using LSDA+U with identical U and J values used in GGA+U calculations. Further, we employed full-potential based WIEN2k code using TB-mBJ functional [32] to substantiate our pseudopotential based calculations. Recently developed TB-mBJ functional [32] has been reported to reproduce the experimental band gap quite accurately in a number of systems [32].

**Results and Discussion:**

**(a) Ellipsometric determination of the optical properties of GaFeO$_3$ thin films**
Fig. 1(a) shows the XRD spectrum of an as-grown GFO film over the 2θ range from 15° to 85° showing (010)-type reflections of orthorhombic structure of GFO (shown in the right inset), indicating epitaxial nature of the film. Absence of any other peaks in the XRD spectra suggests that the film is free of any impurity phase. The out-of-plane lattice parameter (*b* = 9.3973 Å) estimated from the peak positions shows a close agreement with the previously reported XRD data on single crystal GFO (*b* = 9.3950 Å) [23] indicating that the film is fully relaxed along *b*-direction. However, in-plane lattice parameters are strained by ~ 1.63 % due to a mismatch with the substrate lattice parameters. Presence of large strain is also depicted by noticeably large full width half maximum (FWHM) of the rocking curve analysis of (040) peak, as shown in the left inset of Fig. 1(a).

Ellipsometric measurements provide a relative change of the amplitude and phase of linearly polarized monochromatic light reflected from the sample surface, with respect to the incident light. Ellipsometric parameters, $\psi$ and $\Delta$ are related to sample's optical and structural properties by: $\frac{R_p}{R_s} = \tan\psi \cdot e^{i\Delta}$ where, $R_p$ and $R_s$ are the coefficients of reflection of polarized light parallel and perpendicular to the plane of incidence, respectively. [33] Ellipsometric parameters, $\psi$ and $\Delta$ can further be used to describe two intensity parameters termed as:

$$IS = \sin 2\psi \sin \Delta$$
$$IC = \cos 2\psi \cos \Delta \quad (1)$$

In the present work, we have used a three layer model, as shown in the inset of Fig. 1(b), to analyze the ellipsometric data. The layer, labeled as L2 represents the actual film while L1 and L3 take into account the substrate-film interface and film roughness, respectively. The dispersion in L1 consists of 50 % film and 50 % substrate while that in and L3 consists of 50 % film and 50 % void. In order to derive the complex dielectric function and other optical properties from our ellipsometry data, we used Tauc-Lorentz (TL) model [34] in which the imaginary part of the dielectric function is given by:

$$\varepsilon'' = \frac{1}{E} \cdot \frac{AE_0 \Gamma (E-E_g)^2}{(E^2 - E_0^2)^2 + (\Gamma E)^2} \quad for\ E > E_g$$
$$= 0 \quad for\ E \leq E_g \quad (2)$$

where $A$ is amplitude factor which is a function of material density and the momentum matrix element, $E_0$ is peak transition energy corresponding to Penn gap, $\Gamma$ is broadening parameter related to crystallite size [35] and $E_g$ is the band gap energy. To take into account the substrate effect, we assigned a three oscillator Tauc-Lorentz model [34] which described the dispersion of a bare substrate satisfactorily. Fig. 1(b) shows the experimental data (symbols) and corresponding fits (solid lines) of IS and IC of the GFO film. Our ellipsometry data showed a surface roughness of ~ 21 nm and a film thickness of ~ 85 nm which were consistent with our atomic force microscopy (AFM) and surface profilometer measurements. The fitting parameters are listed in the Fig. 1(b)

**(b) Comparison of ellipsometric data with the ground state properties with no external perturbation**

Ellipsometry data along with the results of the density functional calculations of real ($\varepsilon'$) and imaginary ($\varepsilon''$) components of dielectric function are plotted in Fig. 2 (a). Simulation of the ellipsometric data using Tauc-Lorentz (TL) model [34] yielded an energy band gap, $E_g$, of ~ 2.28 ± 0.08 eV, consistent with our ground state electronic structure calculations [14, 24] but lower than the previous experimental data. [16] The plots of $\varepsilon''$ versus photon energy show that absorption in our samples begins at ~ 2-2.5 eV. For a clear visualization of the absorption edge, absorption coefficient ($\alpha$) is plotted semi-logarithmically, as shown in Fig. 2(b). Further, initial part of the absorption spectra beyond the band gap shows a quadratic dependency on incident photon energy (solid line in of Fig. 2(b)) indicating GFO to be an indirect band gap semiconductor. From the fitting of the absorption spectra, the indirect band gap was estimated as $E_g \sim 2.28 \pm 0.02$ eV which is in excellent agreement with the one obtained from the simulation of the ellipsometric data using TL model. However, this is in contrast with our earlier calculations [14] on the ground state structure where we found that direct and indirect band gaps in GFO are identical. We find that below the band edge, the values of $\varepsilon''$ and $k$ are zero while $\varepsilon'$ and refractive index ($n$) possess dispersive behavior, as a function of photon energy, as shown in the inset of Fig.2 (b). To understand the origin of the experimental optical behavior in GFO, we performed first-principles studies using both pseudopotential (GGA+U and LSDA+U) and full-potential (TB-mBJ) based approaches. A comparison, as shown in Fig. 2(a), demonstrates that while our experimental data is consistent with a previous report on single crystal GFO, [25] our LSDA+U, GGA+U and TB-mBJ calculations do not reproduce the experimental data very well. While LSDA+U and GGA+U underestimate the band gap (red shift of the absorption edge with respect to the experiment) TB-mBJ yields a good agreement of the band gap with the experiment. The difference in the intensities between the experimental data and the calculated profiles could be attributed to a number of factors namely, sample quality, temperature, difference between experimental and the ground state crystal and magnetic structures, type of approximation scheme used in the first-principles calculations and the type of broadening used in the experimental and calculated data. [36]

Overall, since our GGA+U profile matched best (among all the calculated results) with the experimental data, subsequent discussions are limited to GGA+U results only. We first calculated the ground state dielectric properties, $\varepsilon'$ and $\varepsilon''$, along the three principal crystallographic directions as plotted in Fig. 2(c). Here, we observe that the optical constants of GFO are anisotropic in nature, a feature consistent with the orthorhombic symmetry of the unit cell and also supported by previous report on GFO single crystal. [25] Further, we also identified the major features (peaks) in the $\varepsilon''$ plot which, in case of insulators like GFO, originate primarily from the inter-band transitions, *i.e.*, from valence (VB) to conduction (CB) bands. We computed the electronic band structure and density of states using GGA+U to identify the transitions responsible for the optical activities in GFO and the results are shown in Fig. 3 (a) and (b). The allowed optical transitions are labeled in the band structure. The density of states plot shows that the upper most part of the VB, -2 eV to 0 eV, is dominated by O 2p states with rather suppressed Fe 3d and Ga 4p states. On the other hand, the lower part of CB is mostly occupied by Fe 3d bands with a subdued presence of O 2p and Ga 4s states. Thus, we can conclude that the optical transitions labeled in Fig. 2(c) and Fig. 3(a) involve transitions from O 2p to Fe 3d states which is further corroborated by the experimental work of Kalashnikova *et al.* [25] Small occupation of O 2p states near the conduction band minimum indicates that GFO has a significant ionic character which is substantiated by our previous charge density and electron localization function (ELF) calculations [14] interpreting negligible ELF values at Fe sites as a signature of complete charge transfer between Fe and O.

O2p-Fe3d electronic transition could further be elucidated using crystal field theory. Within the regular $FeO_6$ octahedral crystal environment with $O_h$ point symmetry, five Fe 3d and 18 O 2p atomic orbitals construct Fe 3d–O 2p bonding and antibonding molecular orbitals (MO), $e_g$ and $t_{2g}$, respectively. In addition, O2p $\pi$ - O2p $\pi$ hybridization leads to oxygen nonbonding orbitals are $t_{1g}(\pi)$, $t_{2u}(\pi)$, $t_{1u}(\pi)$, $t_{1u}(\sigma)$ and $a_{1g}(\sigma)$. [37] The relative energy states of these orbitals play a pivotal role in determining O 2p - Fe 3d transitions responsible for the observed optical activity of GFO. Selection rule allows six transitions in the strong absorption region (energy level ≥ 3.0 eV): $^6A_{1g} \rightarrow {}^6T_{1u}$ related to one electron transition between $t_{2u}(\pi)$, $t_{1u}(\pi)$, $t_{1u}(\sigma)$ and $t_{2g}$ and $e_g$ levels, [38, 39]. However, MOs in the $O_h$ point symmetry further split due to non-cubic ($D_{2h}$) crystal field distortion in GFO and such symmetry lowering in the actual crystal environment would lift some of the restrictions of the transitions leading to the appearance of many more transitions, as shown by several peaks in Fig. 2(c).

**(c) Effect of external perturbations on the ground state optical properties**

The discussion in the previous section was based on the assumption that GFO has antiferromagnetically ordered bulk structure with no site disorder at 0 K. These assumptions are often challenged because materials in thin film forms experience substantial substrate induced strain leading to structural distortion. Consequently, such distortions give rise to modifications in the inter-ionic bond spacing and angles affecting the electronic structure and materials properties. For example, a number of ferroelectric oxides have been reported to demonstrate large variations in the polarization upon application of epitaxial strain. [40] Moreover, experimental structure of GFO is shown to possess significant cation site disorder among Ga and Fe sites driven by their similar ionic sizes [21, 23] which is also ignored in the ground state calculation. In the following sections, we introduce these structural changes in the ground state structure and compare the results of GGA+U calculations with the experiments.

**(i) Effect of epitaxial strain**

First, we analyze effect of epitaxial strain on the optical properties and electronic band structure of GFO. The range of strain chosen is on the basis of present and past experiments [16] where choice of substrate leads to a misfit strain of the order of ~1-3 %. Here, first we plot the real and imaginary parts of dielectric constant ($\varepsilon'$ and $\varepsilon''$) as a function of incident photon energy (Fig. 4(a)) with varying magnitudes of strain. We find that the nature of the $\varepsilon'$ plot remains almost identical to that of ground state structure for strain = ± 1 %. Further $\varepsilon''$ plot shows that the peak at ~ 3.8 eV remains similar for the ground state and -1% strain and it splits into two peaks for +1% strain with splitting further getting pronounced upon increasing the strain to +3%. However, the low energy regions of the $\varepsilon''$ spectra remain identical with no noticeable shift of the absorption edge with the application of strain. These observations indicate that the application of epitaxial strain on the ground state structure alone does not improve the agreement between the experimental and calculated dielectric spectra in GFO.

In addition, the refractive index (at 3 eV) increased linearly with increasing applied strain. To identify the origin of such strain dependent optical activity, we compared the electronic structure obtained at 0% and +3% strains. These showed that the PDOS of Fe 3d states slightly shifted towards higher energy in the conduction band with the application of tensile strain which is attributed to the reduction of some of the Fe-O bond lengths resulting in overlapping wavefunctions and consequent hybridization.

**(ii) Effect of cation site disorder**

So far, our calculations were limited to the ground state antiferromagnetic structure of GFO assuming that there was no cation site disorder. However, the actual structure of GFO always contains cation site disorder driven by quite similar ionic sizes of Ga and Fe. [21, 23] Our previous work demonstrated that site disorder between Fe2 and Ga2 sites is most probable followed by Fe1 and Ga1 sites. [24] We incorporated these Ga-Fe site disorders, one at a time, to study their effect on the optical response in GFO in the unstrained structure. Since the structure of GFO contains four equivalent ions of each cation, exchange of ionic sites between one Fe1/Fe2 to one Ga1/Ga2 would lead to a 25% site disorder in the structure. The degree of disorder, particularly Fe2-Ga2 disorder conceived here is similar to the experimental structure. [21] Calculated dielectric constants of GFO consisting of cation site disorder are plotted in Fig. 4(b) along with those obtained on thin film samples, and the ordered ground state structure. A close inspection of the $\varepsilon''$ spectra near the absorption edge reveals that while Fe1 to Ga1 site interchange does not affect the position of the absorption edge with respect to that of the ground state structure, Fe2 to Ga2 site interchange imparts a leftward shift indicative of a reduction in the band gap. Further, we also find that with Fe1-Ga1 site interchange, the peak in the $\varepsilon''$ plot at ~ 2.86 eV (peak A in Fig. 2(c)) is suppressed whereas it almost vanishes for Fe2-Ga2 interchanged structure. On the other hand, the peak at ~ 2.70 eV in the $\varepsilon'$ spectra of the ordered ground state structure is effectively flattened for Fe2-Ga2 site interchange resulting in a remarkably closer resemblance of the calculated with the experimental data reported by Kalashnikova *et al.* [25] and a much improved match with our experimental results. The observed similarities between the experimental and Fe2-Ga2 site interchanged spectra further substantiate the fact that GFO has inherent cation site disorder with a predominant Fe2-Ga2 site exchange.

Subsequent calculations of electronic band structure and site projected density of states (Fig. 3(c) and (d)) reveal that the evolution of band structure upon incorporating the site disorder differs significantly from that of the ground state band structure. We find that the band gap of site-interchanged GFO is of indirect type ($E_g$ ~ 1.82 eV), consistent with our ellipsometry measurement. The difference in magnitude could be attributed to the GGA method used for the band structure calculation. Site projected DOS plots show that there is a shift of Fe2 3d states towards lower energy which is translated into the down shift of the bands in the band structure and is responsible for the observed red shift of the absorption edge in Fig. 4(b). The reduction of the band gap upon Fe2-Ga2 site interchange is related to the reduction in some of the Fe2-O bond lengths (in the disordered structure) with respect to the corresponding Ga2-O bonds (in the ground state structure) and variation of crystal environment upon imparting the disorder. Reduction of the bond length would induce stronger hybridization and consequently widen the band dispersion leading to a reduction in the band gap.

Disorder induced variation of crystal environment could further be studied by a comparison of electron localization function (ELF) of the ground state structure and the Fe2-Ga2 site interchanged structure (not shown here). While finite ELF values between Ga1-O and Ga2-O indicate significant covalency, complete charge transfer between Fe2 and O sites is evident from zero ELF value at Fe2 site and across Fe2-O bonds. Thus site disorder modifies the ELF mapping within the unit cell and consequently the crystal environment. Such variation of the crystal environment is believed to be responsible for the evolution of electronic structure and consequent optical spectra in the disordered GFO.

**(iii) Combined effects of site-disorder and epitaxial strain**
Finally, to completely mimic the experimental scenario in GFO epitaxial thin films, we applied tensile epitaxial strain to Fe2-Ga2 site disordered structure (as discussed in section (ii)) and calculated the optical properties and then compared the results with our experimental data of thin films (Fig. 4(c)). Here we observe that the agreement between the experiment and the calculations further improves significantly upon application of tensile strain on the structure with 25% Fe2-Ga2 cation site disorder. While the band gap calculations suffer from inherent limitation of LDA and GGA techniques and hence are unreliable, the calculated spectra obtained upon applying tensile strain on Fe2-Ga2 site disordered GFO is remarkably similar to the experimentally obtained dielectric function as shown in Fig. 4(c). The difference in the absolute intensities can be attributed to various extrinsic parameters such as the sample quality, temperature etc. The electronic structure of the modified structure of GFO would have contributions from the above two effects described in 3(b) and 3(c) where disorder drives the Fe 3d bands of the Fe2 ion at the Ga2 site to shift downward while tensile strain alters the cation-oxygen bond-lengths and angles which in turn alter the positions of the electronic states. We further compared the experimental optical constants with these calculations over the experimental measurement domain as shown in Fig. 5. In Fig 5(a), we show the calculated and experimental results for $n$ and $k$ which match well with each other. It is found that experimentally determined $n$ starts with a finite value ($n \sim 2.1$) and then slowly increases with energy showing a peak ($n \sim 2.5$) at $\sim 3$ eV. On the other hand, the calculated spectra show the position of this peak at slightly lower energy due to underestimation of the band gap by GGA+U method. The peak in $n$ spectra can be attributed to the beginning of absorption in the material triggered by transition from valence band O 2p to conduction band Fe 3d state. On the other hand, the result for $k$ starts with zero value and then begins to increase at energy corresponding to the band gap. Since our calculated gap is lower compared to the experimental gap, the experimental curve starts to rise at higher energy. However, calculated spectra could be appropriately scaled (shifted by a constant energy), to yield minimum qualitative mismatch of the overall profile between calculated and experimental spectra. Magnitudes of $n$ and $k$ show a clear departure for the experiment and calculations, attributed to the factors such as sample quality, temperature etc. as mentioned before. Reflectivity ($R$) spectra in Fig. 5(b), shows an initial gradual increase followed by a peak ($R \sim 0.19$) near the absorption edge which is consistent with the absorption spectra, also shown in Fig. 5(b). Again, the difference between the calculated and the experimental $R$ spectra could be attributed to the band gap underestimation by GGA+U calculations. Optical conductivity, as shown in Fig. 5(c), demonstrates an onset above 2 eV for both experiment ($\sim 2.5$ eV) and calculation ($\sim 2.1$ eV) and level off to values, 7500 $\Omega^{-1}$cm$^{-1}$ and 13000 $\Omega^{-1}$cm$^{-1}$ at 4 eV for experimental and calculated spectra, respectively.

**(iv) Effect of off-stoichiometry**
Since GFO has large compositional tolerance, [22] it would be of interest to examine the effect of off-stoichiometric Ga:Fe ratio on the optical properties and consequent changes in the electronic structures. This is important because it results in an imbalance in the cation site occupation between Fe and Ga and its effects are manifested in the magnetic behavior of GFO [22]. In this context, we have considered two cases: first, considering substitution of one Ga ion at Ga2 site by one Fe ion at Fe2 site and second, considering substitution of two Ga ions at Ga2 sites by two Fe2 site ions. These scenarios resemble two Fe-excess compositions $x = 1.125$ and $x = 1.25$ in $Ga_{2-x}Fe_xO_3$, well within the experimentally obtained single phase domain of GFO. [22] Subsequently, we relaxed the two structures and computed the optical properties and the electronic structures. Here, we have not included the site disorder allowing us to exclusively investigate the effect of off-stoichiometry.

It was found that the electronic band structure and the density of states (plots not shown) calculations of these off-stoichiometric compositions show similar effects as observed for the disordered GFO. It was observed that with increasing Fe content, Fe ions substituting Ga2 sites would have Fe 3d states at increasingly low energies resulting in a monotonic decrease in the band gap. Lowering of crystal symmetry due to doping further induces band splitting of Fe 3d band in these cases. Resulting modifications in the electronic structures thus, affect the optical spectra for compositions having excess Fe content. Fig. 6(a) shows the yy component of real and imaginary parts of dielectric constant tensors with different stoichiometry *viz.*, $x = 1.0$, $x = 1.125$ and $x = 1.25$. We observe that with increasing Fe content the fundamental absorption edge shifts towards lower energy consistent with the previous experimental observations on single crystal [25] and our band structure calculations which show a reduction in the band gap with increasing Fe content ($x$). Moreover, the intensity of dielectric function increases with increasing Fe content.

**(v) Effect of hydrostatic pressure**
Finally, we take into account the structural distortion which can induce the instability leading to phase transformation in a few systems with significant magneto-structural coupling. [41, 42] Such distortion which can be brought about by applying hydrostatic stress has also been reported to alter the magnetic behavior in GFO owing to the presence of magneto-structural coupling. [24] Here, we study the evolution of optical constants of GFO as a function of distortion induced by hydrostatic stress. The evolution of yy component of real and imaginary parts of dielectric constants as a function of application of hydrostatic pressure is shown in Fig. 6(b). The figure shows that with increasing hydrostatic pressure, the position of the first peak in the $\varepsilon''$ spectra remains almost identical, while the peak at 3.78 eV tends to shift towards higher energy. An additional peak also appears which is marked with the vertical arrow beyond 20 GPa. The evolution of such optical behavior has its origin in the electronic structure as we explain below. With increasing hydrostatic pressure, since the structure is distorted as the bond lengths, in general, are decreased. As a result, there is a growing tendency of the wave functions of the adjacent ions to overlap with each other. Consequently, the energy levels shift in a repulsive manner which is reflected in the observed optical behavior. Such modification of electronic structure of GFO upon application of pressure has been explained in our previous work. [24] The inset of Fig. 6 (b) shows yy component of the refractive index ($n$) as a function of applied pressure exhibiting a gradual fall at the incident photon energy of 3 eV. Thus, this study qualitatively demonstrates the presence of a coupling between the optical and the structural parameters.

**IV. Conclusions**

In summary, we have performed ellipsometry studies on epitaxial GaFeO$_3$ thin films and have compared the dielectric response and other optical constants with our density functional calculations using different approximation schemes with GGA+U showing the best agreement with the experiments. The origin of optical activities in GFO is identified as transition from O 2p to Fe 3d states. We find that the inclusion of site disorder, off-stoichiometry, epitaxial strain and hydrostatic pressure influence the optical properties due to shifting of Fe 3d state. We observe that incorporation of the cation site disorder into GFO lattice renders it to become an indirect band gap semiconductor, consistent with the experimental observations. Further, the cation site disorder also brings about a significant reduction in the electronic band gap with respect to that of the ground state structure of GFO. Interestingly, we find that inclusion of site disorder and epitaxial strain into the ground state structure significantly improves the agreement between calculated and experimental results clearly illustrating that gallium ferrite contains inherent cationic site-disorder.

**Acknowledgements**
The work was supported by Department of Science and Technology, Govt. of India through project number SR/S2/CMP-0098/2010. Authors thank Prof. Y.N. Mohapatra, SA thanks NPL for the J C Bose Fellowship.

# **Figure Captions**

Fig. 1 (a) XRD spectrum of a pulsed laser deposited GFO thin film showing (010) orientation (the left inset shows the rocking curve of (040) peak and the right inset shows a schematic of the orthorhombic unit cell of GFO); (b) fitting of ellipsometry data using Tauc-Lorentz model with fit parameters listed inside the plot (inset shows the three layer model used for simulation).

Fig.2 (a) Real ($\varepsilon'$) and imaginary ($\varepsilon''$) parts of dielectric function determined experimentally and theoretically and compared with the literature; (b) experimentally determined absorption coefficient ($\alpha$) showing the absorption edge (inset plots dispersion of experimentally computed refractive index ($n$) and extinction coefficient ($k$)) and (c) $\varepsilon'$, $\varepsilon''$ spectra along principal crystallographic directions corresponding to the ground state structure of GFO calculated using GGA+U method, plotted as a function of incident photon energy

Fig.3 (a) Electronic band structure of the ground state structure of GFO with arrows indicating the interband transitions responsible for the evolution of the peaks in the $\varepsilon''$ spectra shown in Fig.1 (c); (b) total and partial density of states of ground state structure of GFO; (c) comparison of band structures of the ground state and Fe2-Ga2 site disorder structures and (d) total and partial density of states of Fe2-Ga2 site disordered structure.

Fig.4 Real ($\varepsilon'$) and imaginary ($\varepsilon''$) components of dielectric function plotted as a function of incident photon energy obtained experimentally and theoretically showing effect of (a) epitaxial strain, (b) site disorder and (c) epitaxial strain and Fe2-Ga2 site disorder.

Fig.5 Comparison of experimentally determined optical constants with Fe2-Ga2 site interchanged structure with tensile strain of 3 %.

Fig.6 Effect of (a) off-stoichiometry and (b) hydrostatic pressure on the $\varepsilon'$, $\varepsilon''$ spectra of GFO, inset shows the variation of refractive index at 3 eV as a function of applied hydrostatic pressure.

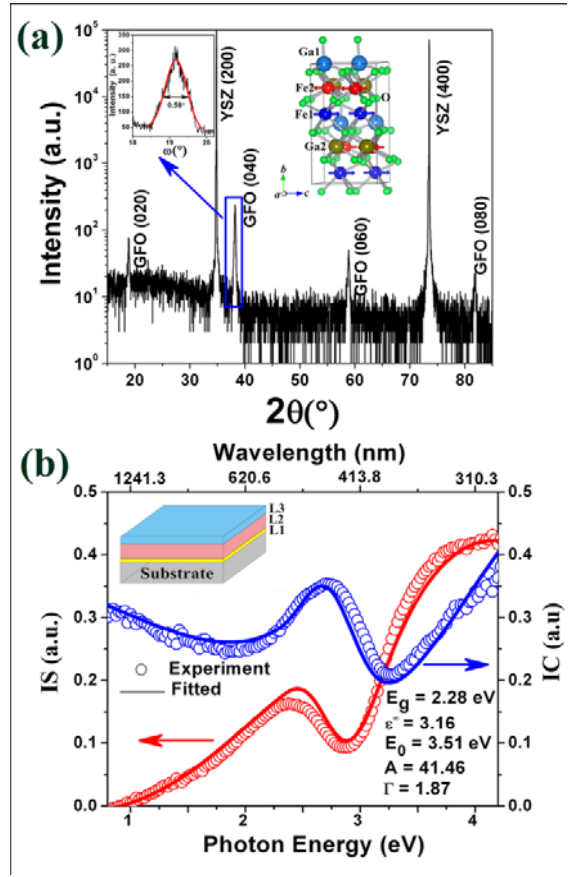

Fig. 1 Roy et al

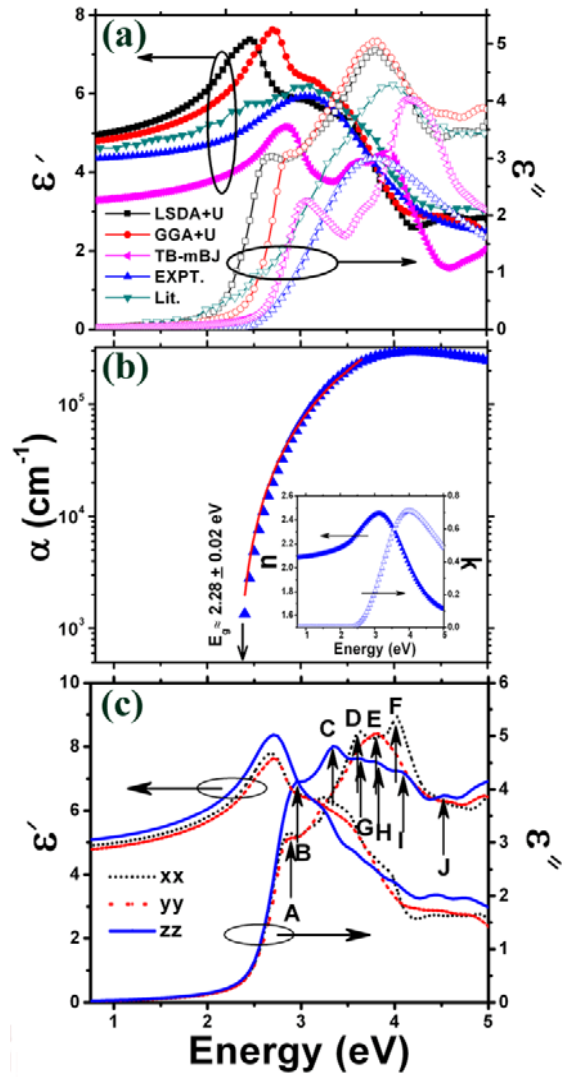

Fig. 2 Roy et al

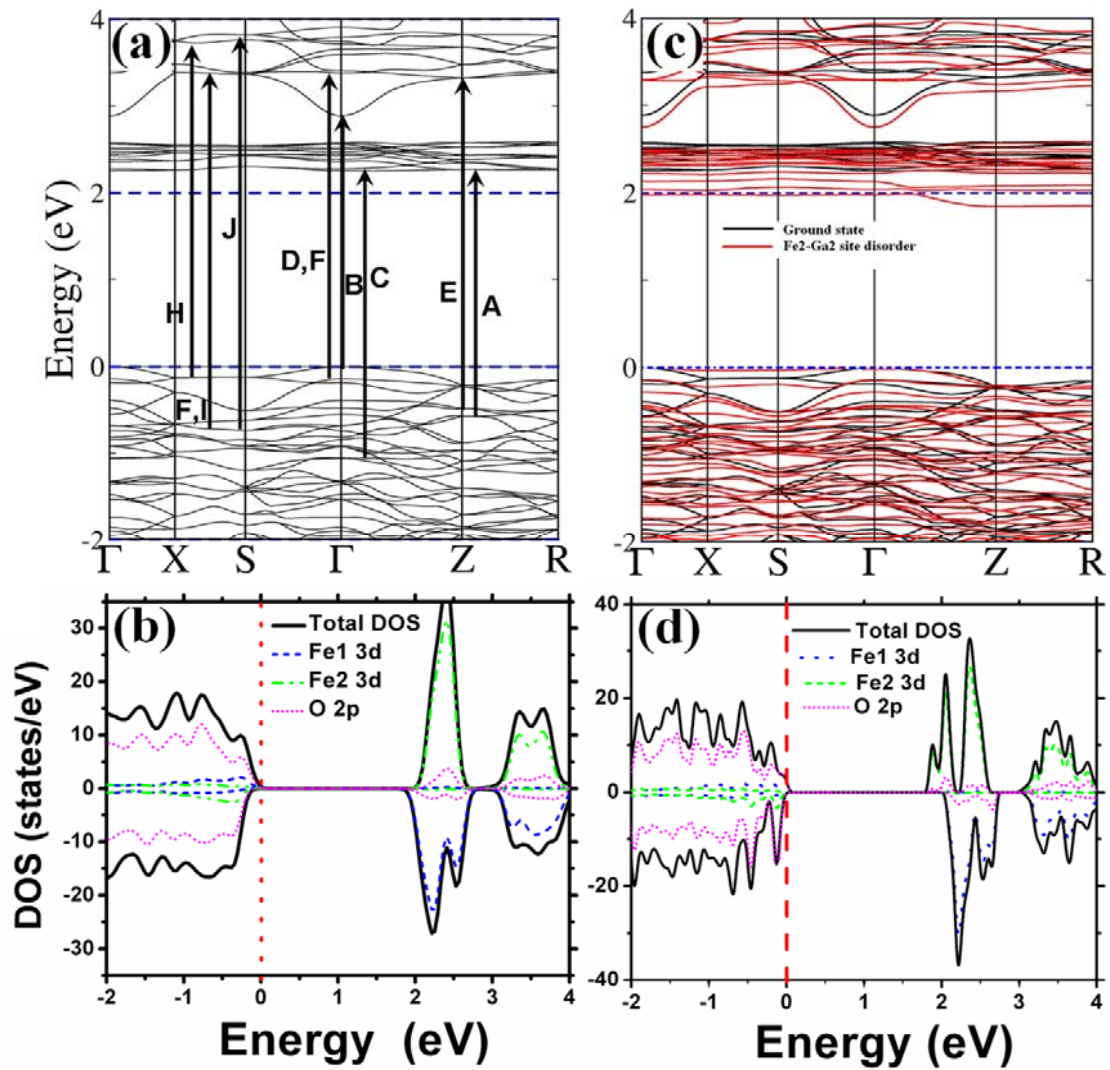

Fig.3 Roy et al

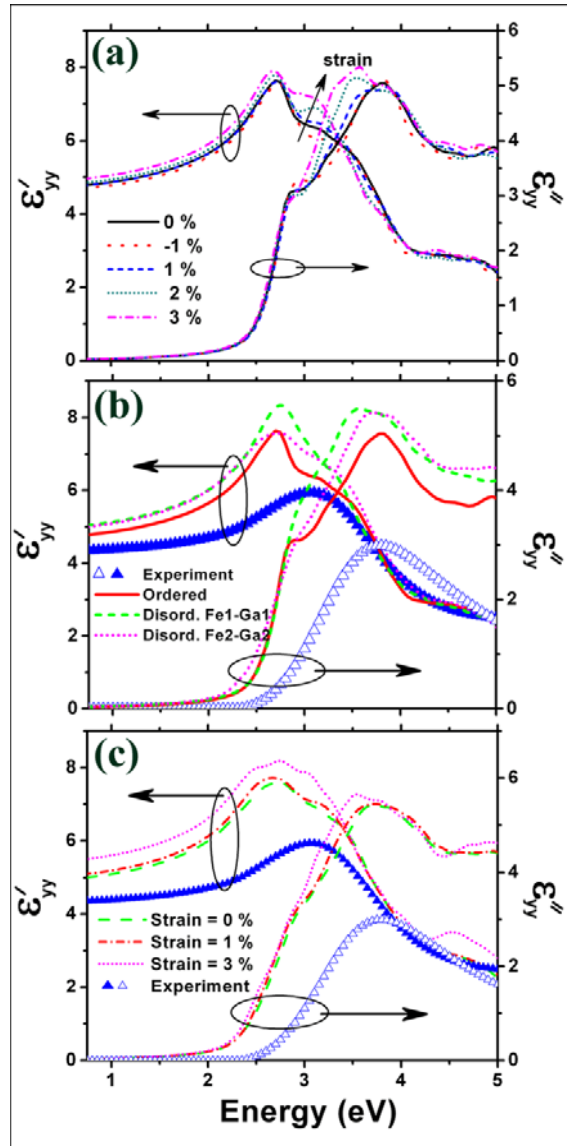

Fig.4 Roy et al

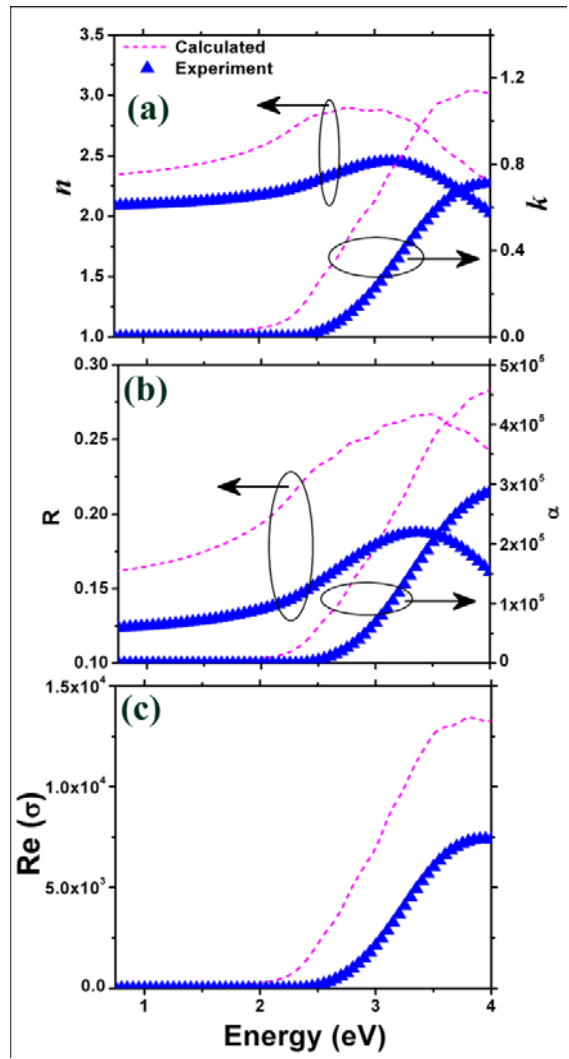

2
3
4   Fig.5 Roy et al
5

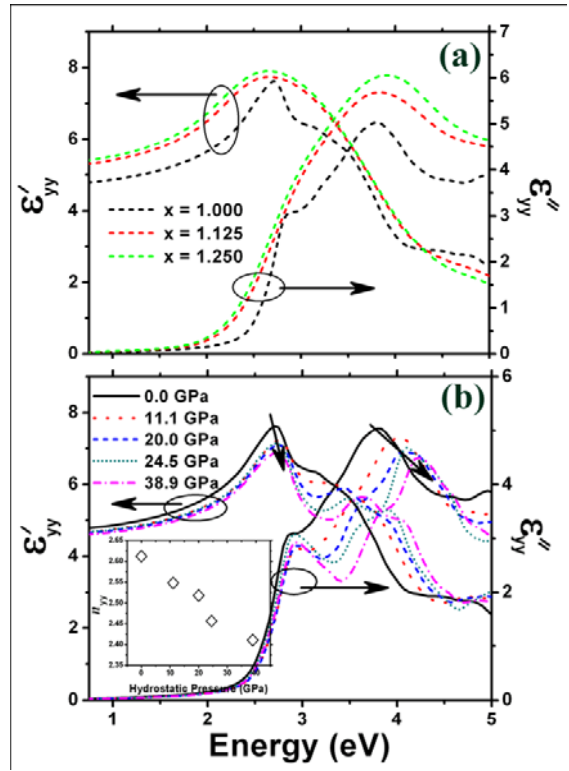

Fig. 6 Roy et al